\documentstyle[12pt]{article}
\newcommand{\be}{\begin{equation}}
\newcommand{\ee}{\end{equation}}
\newcommand{\ba}{\begin{eqnarray}}
\newcommand{\ea}{\end{eqnarray}}
\newcommand{\baa}{\begin{eqnarray*}}
\newcommand{\eaa}{\end{eqnarray*}}
\newcommand{\bb}{\begin{thebibliography}}
\newcommand{\eb}{\end{thebibliography}}
\newcommand{\ci}[1]{\cite{#1}}

\newcommand{\lab}[1]{\label{#1}}
\newcommand{\re}[1]{(\ref{#1})}

\newcommand\fac[2]{\mbox{$\frac{#1}{#2}$}}

\newcommand\nal{|n;\alpha\rangle}
\newcommand\nam{|n-1;\alpha\rangle}
\newcommand\nap{|n+1;\alpha\rangle}
\newcommand\om{\omega}
\newcommand\Om{\Omega}
\newcommand\rar{\rightarrow}
\newcommand\lbr{\lbrace}
\newcommand\rbr{\rbrace}
\newcommand\ka{\kappa}

\newcommand\al{\alpha}
\newcommand\ga{\gamma}
\newcommand\de{\delta}
\newcommand\lphi{\langle\phi_s|}
\newcommand\rpsi{\psi_p\rangle}

\begin{document}
\bibliographystyle{unsrt}
\begin{titlepage}

{\hfill{ UdeM-LPN-TH-93-146}

{\hfill April 1993}}

\vspace*{25mm}

\begin{center}
{\LARGE \bf Hidden Symmetry of the Racah and Clebsch-Gordan\\[2mm]


Problems for the Quantum Algebra $sl_q(2)$}\\[5mm]


Ya.I.Granovskii \\

{\it Physics Department, Donetsk University, Donetsk, 340055 Ukraine}\\[3mm]

A.S.Zhedanov%
\footnote{On leave of absence from the Physics Department, Donetsk
University, Donetsk, 340055 Ukraine}
\\

\vspace{2mm}

{\it Laboratoire de Physique Nucl\'eaire, Universit\'e de Montr\'eal,
C.P. 6128, \\
Succ. A, Montr\'eal, Qu\'ebec, H3C 3J7, Canada}\\[5mm]
\end{center}

\vspace{5mm}

\begin{abstract}
The Askey-Wilson algebra $AW(3)$ with three generators is shown to serve
as a hidden symmetry algebra underlying the Racah and (new) generalized
Clebsch-Gordan problems for the quantum algebra $sl_q(2)$. On the base of this
hidden symmetry a simple method to calculate corresponding coefficients
in terms of the Askey-Wilson polynomials is proposed.

\end{abstract}
\end{titlepage}

\newpage

\section{Introduction}

\vspace{8mm}

As is well known, the quantum algebra $sl_q(2)$ possesses many remarkable
properties closely related to those of the ordinary $sl(2)$ Lie algebra.
In particular, the Clebsch-Gordan and Racah problems for $su_q(2)$ and
 $su_q(1,1)$ algebras can be formulated and resolved in such
a manner that the corresponding coefficients are expressed in terms of
 the Askey-Wilson polynomials [1-7]. What is
the origin of this "experimental" result? In other words, why can these
coefficients be calculated explicitly in terms of q-orthogonal polynomials?
For the ordinary $su(2)$ and $su(1,1)$ algebras the answer was found in
Ref. \ci{grz1} where the quadratic Racah algebra $QR(3)$ with three generators
was shown to serve as a hidden symmetry algebra underlying the corresponding
Racah and Clebsch-Gordan problems. By the way, this result explains why
the same Racah polynomials arise as $6j$ symbols for such strictly
different algebras as (compact) $su(2)$ and (non-compact) $su(1,1)$.

It is worth mentioning that the representation theory for the quadratic
Racah algebra $QR(3)$ is very simple and can be constructed independently
of a concrete realization in terms of $su(2)$ or $su(1,1)$ generators. So,
really
we are dealing with  an extra-symmetry of some non-linear combinations of
Lie algebra
generators (namely, intermediate Casimir operators - for details see \ci{grz1})
of the added Lie algebras. It is interesting to note that it was Racah who
first introduce in such a manner some non-linear (cubic) algebra in order
to study the representations of the $SU(3)$ group \ci{rac}. In fact the
structure of the algebra introduced by Racah is closely related (up to some
additional term) to that of $QR(3)$. This justifies the name of this algebra.

The purpose of this paper is to present an analogous algebraic treatment
of the Racah and Clebsch-Gordan problems associated with the quantum algebra
$sl_q(2)$.
The paper is organized as follows. In Sec.II, we recall the addition rule
for different types of $sl_q(2)$ algebras in accordance with Ref.\ci{grkh,zh1}.
In Sec.III, the AW(3) algebra is shown to be the hidden symmetry algebra for
the
Racah problem of $sl_q(2)$; this allows one to express all types of
Racah coefficients in terms of Askey-Wilson polynomials. In Sec.IV we show
how the generalized Clebsch-Gordan problem for $sl_q(2)$ can be obtained from
the Racah one by a simple contraction procedure having no any classical
analogue.
This procedure allows one to obtain the corresponding hidden  symmetry
algebra and the
explicit expression for the Clebsch-Gordan coefficients directly
from the Racah case.

\section{Different types of $sl_q(2)$ and their addition rule}

We adopt the standard notation for the $sl_q(2)$ algebra \ci{lisk,grkh,zh1,maj}
\ba
[A_0,A_\pm]&=&\pm A_\pm,
\nonumber \\
\; [A_-,A_+]&=&uq^{-A_0} + vq^{A_0},
\lab{def} \ea
where $q=exp(-\omega), \omega>0$.

We are restricted ouselves to the real forms \ci{lisk} of $sl_q(2)$, i.e. the
parameters $u$ and $v$ are assumed to be real. Moreover, we shall assume
that some unitary representation of $sl_q(2)$ is chosen where the operators
$A_-$ and $A_+$ are Hermitian conjugated whereas the operator $A_0$ is
Hermitian. In what follows we shall denote the algebra $sl_q(2)$ with
commutation relations \re{def}  by the symbol $(u,v)$.

The special cases of $(u,v)$ algebra are:

(i) $su_q(2)$ if $u=-v<0$;

(ii) $su_q(1,1)$ if $u=-v>0$;

(iii) $cu_q(2)$ if $u=v>0$;

(iv) $eu_q^+$ if $u>0, v=0$;

(v) $eu_q^-$ if $u=0, v>0.$

All these case were intensively studied in the literature. Note, that the cases
(iv) and (v) describing two types of q-oscillator algebra can be obtained
from the cases (i)-(iii) by a simple contraction procedure \ci{maj}. It is
worth mentioning that the cases (i)-(iii) describe three different types
of $sl_q(2)$ algebra, i.e. the unitary representations of these can not be
obtained from one another by simple analytic continuation and renormalization
of the initial generators $A_0, A_\pm$ (for example, the $su_q(2)$ algebra
has only finite-dimensional representations, whereas the algebras $su_q(1,1)$
and $cu_q(2)$ have only infinite-dimensional representations - see below).

The Casimir operator of the $(u,v)$ algebra has the expression
\be
\hat \kappa =A_+A_- + (vq^{A_0}-uq^{1-A_0})/(1-q)
\lab{cas} \ee
Fixing the value of the Casimir operator
\be
\kappa(\alpha)=(vq^\alpha - uq^{1-\alpha})/(1-q),
\lab{vcas} \ee
where $\alpha$ is some parameter, we get a unitary representation of the
{\it (u,v)}
algebra. In this paper we restricted ourselves to the representations of the
positive
discrete series $D^+_\alpha$ in some canonical basis $\nal$:
\ba
A_0\nal &=& (\alpha + n)\nal,\qquad n = 0,1,2,...\\
\nonumber\\
A_-\nal &=& r_n\nam,\\
\nonumber\\
A_+\nal &=& r_{n+1}\nap,
\lab{rep} \ea
where
\be
r_n^2=(1-q^n)(vq^\alpha + uq^{1-n-\alpha})/(1-q)
\lab{r}\ee
Note that because $r_0=0$ the state $|0;\alpha\rangle$ is the vacuum of
the representation $D^+_\alpha$.

It is seen from \re{r} that the representation $D^+_\alpha$ exists provided
that
\be
vq^\alpha+uq^{1-n-\alpha} > 0
\lab{cond}
\ee
for all values of $n$. The condition \re{cond} is fulfilled for $su_q(1,1)
(\alpha>0),cu_q(2),eu_q^\pm (\alpha$ is arbitrary real parameter) algebras.
Otherwise, if \re{cond} is fulfilled for $n\leq N$ but
\be
vq^\alpha +uq^{-N-\alpha}=0,
\lab{fin} \ee
then one obtains a $N+1$ dimensional representation (this takes place, e.g.
for $su_q(2))$.

The $(u,v)$ algebra possesses an addition property that allows to
add different types of $sl_q(2)$ algebras \ci{grkh,zh1,maj}:
\ba
A_0^{(3)}&=&A_0^{(1)}+A_0^{(2)},
\nonumber \\
A_\pm^{(3)}&=&A_\pm^{(1)}\exp(\omega A_0^{(2)})+
A_\pm^{(2)}\exp(-\omega A_0^{(1)})
\lab{add} \ea

The addition rule \re{add} is the same as for ordinary $sl_q(2)$ algebra
\ci{maj}, however the algebras $(u_1,v_1)$ and $(u_2,v_2)$ in \re{add} may
have different types. It is easily seen that in order for the
operators $A_0^{(3)}, A_\pm^{(3)}$ to form new $(u_3,v_3)$ algebra,
the following relations must
be fulfilled:
\be
u_3=u_1, \qquad v_3=v_2, \qquad u_2=-v_1
\lab{connect} \ee

In symbolic form the addition rule \re{add} can be written as
\be
(u,v)_1\oplus (-v,w)_2=(u,w)_3
\lab{symb} \ee

It is worth mentioning that this addition rule does not destroy the unitarity
of the representation, i.e. the operator $A_0^{(3)}$ is Hermitian and the
operators $A_-^{(3)}$ and $A_+^{(3)}$ are Hermitian conjugated.
So, if the representations $D^+_{\alpha_1}$ and $D^+_{\alpha_2}$ are given,
then one can  construct the Clebsch-Gordan decomposition
\be
|n_3;\alpha_3\rangle = \sum_{n_1,n_2} (n_1\alpha_1 n_2\alpha_2;n_3\alpha_3)
|n_1;\alpha_1\rangle\otimes |n_2;\alpha_2\rangle,
\lab{Cl} \ee
where the symbol $(n_1\alpha_1 n_2\alpha_2;n_3\alpha_3)$ stands for the
Clebsch-Gordan coefficients (CGC). Obviously, in \re{Cl} the relation
\be
n_1+n_2=n_3+\alpha_3 -\alpha_1 -\alpha_2
\lab{proj} \ee
is sutisfied.

The reciprocal decomposition has the form
\be
|n_1;\alpha_1\rangle\otimes|n_2;\alpha_2\rangle=
\sum_{\alpha_3 =\alpha_1 +\alpha_2}^{\alpha_1 +\alpha_2 +n_1 +n_2}
(n_1\alpha_1 n_2\alpha_2;n_3\alpha_3) |n_3;\alpha_3\rangle.
\lab{rec} \ee

The explicit expression of these CGC (in the case of the representations of
discrete positive series) in terms of Hahn polynomials was found in \ci{kl1,
koo1,lisk} for $su_q(2)$ and $su_q(1,1)$ algebras and in \ci{grkh} for the
case when the algebras of different types are added (note that in \ci{grkh}
addition rule \re{add} was presented in another - but equivalent -
form).

It is worth mentioning  that among the addition rules \re{add} there exists
some
that can not be obtained from well-known addition rules for $su_q(2)$ or
$su_q(1,1)$ by any contraction (with real parameters).

As an example we present some non-trivial addition
\be
(u,0)_1 \oplus(0,u)_2 =(u,u)_3, \qquad u>0,
\lab{exa} \ee
mapping two different q-oscillator algebras $eu_q^+(2)$ and $eu_q^-(2)$ onto
the third q-oscillator algebra $cu_q(2)$. Obviously, such an addition  rule
can not be obtained form $su_q(1,1)$ or $su_q(2)$ cases by means of ordinary
contraction (obviously, contraction with complex parameters destroys the
unitarity of the representations). Note, that the addition rule \re{exa}
allows one to solve the problem of finding the addition rules for the
q-oscillator algebras: there is no such addition involving {\it the same}
q-oscillator algebras, instead, all the added and resulting algebras should be
{\it different} (see also \ci{grkh,yan}).

Of course, there are many other special types of addition rule
having no any classical analogs: for example one can obtain from \re{symb}
a new (non-Schwinger) realization of $su_q(2)$ and $su_(1,1)$ algebras in
terms of two q-oscillator; this realization takes place only in a q-domain
and disappears in the classical limit $q \to 1$ (for details see \ci{zh2}).

\vspace{6mm}
\section{$AW(3)$ algebra and the Racah problem}
The addition rule \re{add} possesses an associativity property when {\it three}
algebras are added.
Indeed, let us have three (mutually commuting) sets of the $sl_q(2)$
generators $A_0^{(i)}, A_\pm^{(i)}, i=1,2,3$ with corresponding indicators
$(u_i,v_i)_i$. If the relations
\be
v_1=-u_2=v, v_2=-u_3=w, u_1=u, v_3=z,
\lab{uvz} \ee
hold, then one can obtain the same resulting algebra $(u,z)_4$ in
two different ways:

i) one can first add the algebras $i=1,2$ obtaining the algebra
\be
(u,w)_{12} = (u,v)_1\oplus (-v,w)_2,
\lab{12} \ee
and then getting the complete sum by adding $i=3$ algebra:
\be
(u,z)_4 = (u,w)_{12}\oplus (-w,z)_3;
\lab{l2,3} \ee

ii) or,one can first add the $i=2,3$ algebras
\be
(-v,z)_{23} = (-v,w)_2\oplus (-w,z)_3,
\lab{23} \ee
to get the complete sum by adding the $i=1$ algebra:
\be
(u,z)_4 = (u,v)_1\oplus (-v,z)_{23}.
\lab{1,23} \ee

According to the schemes i) and ii) we can introduce two intermediate Casimir
operators $\hat \kappa_{12}$ and $\hat \kappa_{23}$ corresponding to the
$(u,w)_{12}$ and $(-v,z)_{23}$ algebras:
\ba
K_1= \hat \kappa_{12} ={A_+^{(1)}A_-^{(2)}\exp(\om (A_0^{(2)}-A_0^{(1)}-1)) +
h.c.} + \kappa_1 \exp(2\om A_0^{(2)})
\nonumber \\
+ \kappa_2 \exp(-2\om A_0^{(1)}) - v\coth\om \exp(2\om(A_0^{(2)}-A_0^{(3)})),
\lab{K1} \\
K_2= \hat \kappa_{23} ={A_+^{(2)}A_-^{(3)}\exp(\om (A_0^{(3)}-A_0^{(2)}-1)) +
h.c.} + \kappa_2 \exp(2\om A_0^{(3)})
\nonumber \\
+ \kappa_3 \exp(-2\om A_0^{(2)}) - w\coth\om \exp(2\om (A_0^{(3)}-A_0^{(2)})),
\lab{K2} \ea
where $\kappa_i  , (i=1,2,3)$ stands for the values of corresponding Casimir
operators $\hat \kappa_i$ obviously commuting with $K_1$ and $K_2$.

The full Casimir operator
\ba
\hat \kappa_4 &=& A_+^{(4)}A_-^{(4)} + (z\exp(-2\om A_0^{(4)}) - u\exp
(2\om (A_0^{(4)}-1))/(1-q) \\
\nonumber
&=&
\quad K_1\exp(2\om A_0^{(3)}) + K_2\exp(-2\om A_0^{(1)}) - \kappa_2 \exp
(2\om (A_0^{(3)}-A_0^{(1)})) \\
\nonumber
&+&[A_-^{(1)}A_+^{(3)}\exp(\om (A_0^{(3)}-A_0^{(1)}-1)) + h.c.]
\lab{full} \ea
also commute with both $K_1$ and $K_2$ and can be replaced by the constant
$\kappa_4$.

The Racah problem consists in finding the overlaps between eigenstates of the
intermediate Casimir operators $K_1$ and $K_2$ in the space with fixed
values of $\kappa_i$,  $i=1,2,3,4$. This problem is non-trivial because
the operators $K_1$ and $K_2$  do  not commute with one another.

The crucial observation in our considerations is that the operators $K_1$ and
$K_2$ are closed in frames of simple algebra with three generators.

In order to see this let us introduce the procedure of "q-mutation" for
arbitrary operators $L,M$
\be
[L,M]_{\om}\equiv e^{\om} LM - e^{-\om} ML
\lab{qmut} \ee

A direct calculation shows that operators $K_1,K_2$ together with
their q-mutator $K_3$ obey the following algebra
\ba
[K_1,K_2]_{\om}&=&K_3,
\nonumber \\
\quad[K_2,K_3]_{\om}&=&BK_2+C_1K_1+D_1,
\nonumber \\
\quad[K_3,K_1]_{\om}&=&BK_1+C_2K_2+D_2,
\lab{AW3} \ea
where $B,C_{1,2},D_{1,2}$ are the structure constants of the algebra \re{AW3}:
\ba
B&=&4\sinh^2\om (\ka_1 \ka_3 + \ka_2 \ka_4),
\nonumber \\
C_1 &=& 4vz\cosh^2\om,\qquad C_2 = -4uw\cosh^2\om,
\nonumber \\
D_1 &=& -2\sinh2\om (z\ka_1 \ka_2 + v\ka_3 \ka_4),\quad D_2 = 2\sinh2\om
(u\ka_2 \ka_3 - w\ka_1 \ka_4).
\lab{str} \ea \\
The algebra \re{AW3} is known as the Askey-Wilson algebra with three generators
$AW(3)$. It was introduced and  studied in \ci{zh3,ann}. The Casimir
operator $\hat Q$
commuting with all the generators $K_1,K_2,K_3$ of the $AW(3)$ algebra
has the expression
\ba
\hat Q= \fac 12 \lbr K_3,\tilde K_3\rbr  + \cosh2\om (C_1K_1^2 +
C_2K_2^2)
\nonumber \\
\quad + B \lbr K_1,K_2\rbr + 2\cosh^2\om (D_1K_1+D_2K_2),
\lab{AWQ} \ea
where the symbol $\lbr.,.\rbr$ stands for the anticommutator and
$\tilde K_3$ is the
"dual" generator:
\be
\quad \tilde K_3 = [K_1,K_2]_{-\om}=e^{-\om}K_1K_2-e^{\om}K_2K_1
\lab{dual} \ee \\
Given the realization \re{K1}, \re{K2} of the $AW(3)$ algebra the Casimir takes
the value
\ba
Q= 4[-uvwz\cosh^4\om \sinh^{-2}\om + \sinh^2\om (\ka_1^2\ka_3^2 +\ka_2^2
\ka_4^2)-2\sinh^2\om \cosh2\om \ka_1\ka_2\ka_3\ka_4
\nonumber \\
 + \cosh^2\om (-wz\ka_1^2 +uz\ka_2^2 +uv\ka_3^2 -vw\ka_4^2)].
\lab{Qval} \ea \\
So, the operators $\hat\ka_{12}$ , $\hat\ka_{23}$ together with their
q-mutator form
a realization of the $AW(3)$ algebra with fixed values of the structure
constants \re{str} and Casimir operator \re{Qval}.

So far, we have not made any assumptions on the concrete type of the
$(u_i,v_i)_i$
representations. Now let us assume that all the added algebras $(i=1,2,3)$
belong to the positive discrete series $D_{\alpha_i}^+$
\re{rep} (the finite-dimensional case \re{fin} is also admitted). Then the
resulting algebra $i=4$ has also the representation of $D_{\alpha_4}^+$.
It is easily seen that on the space with fixed values $\alpha_i, i=1,2,3,4$
the operators $K_1$ and $K_2$ become finite-dimensional matrices:
\ba
K_1\psi_p = \ka_{12}(p)\psi_p,
\lab{K1s} \\
K_2\phi_s = \ka_{23}(s)\phi_s,
\lab{K2s} \ea
where $\psi_p$ and $\phi_s$ are some eigenstates of the operators $K_1$ and
$K_2$; the corresponding eigenvalues are
\ba
\ka_{12}(p) &=& (wq^p -uq^{1-p})/(1-q),
\lab{ka12} \\
\ka_{23}(s) &=& (zq^s + vq^{1-s})/(1-q)
\lab{ka23} \ea

The discrete parameters $p$ and $s$ take the values:
\ba
p=\al_{12} = \al_1+\al_2, \al_1+\al_2+1,...,\al_4-\al_3,
\nonumber \\
s =\al_{23}= \al_2+\al_3, \al_2+\al_3+1,...,\al_4-\al_1.
\lab{ps} \ea

It is clear that
\be
p_{max}-p_{min}=s_{max}-s_{min}=N,
\lab{fd} \ee
where
\be
N=\al_4-\al_1-\al_2-\al_3=0,1,2,...
\lab{N=} \ee
so $N+1$ is the dimension of the space where the operators $K_1$ and $K_2$ act,
in other words, $N+1$ is the dimension of $AW(3)$ representation (strictly
speaking, the relation \re{N=} is valid when all the representations
$D_{\al_i}^+$ are infinite-dimensional; if some of these are
finite-dimensional (i.e. for
$su_q(2)$) then apart from \re{N=} there are another possibilities; we shall
assume, however, that the relation \re{N=} is fulfilled also in this case).

As was shown in \ci{zh3}, all finite-dimensional representations of $AW(3)$
algebra are easily obtained and have the following important properties:

(i) If $\psi_p$ are the eigenstates of the operator $K_1$
\be
K_1\psi_p=\lambda_p\psi_p,
\lab{eK1} \ee
then the operator $K_2$ is three-diagonal in this basis:
\be
K_2\psi_p = a_{p+1}\psi_{p+1}+a_p\psi_{p-1}+b_p\psi_p,
\lab{thr} \ee
where the spectrum $\lambda_p$ and the matrix coefficients of the
representation $a_p,b_p$ are expresssed by the formulae:
\ba
\lambda_p &=& C_2q^{-p}+q^p/(q-q^{-1})^2,
\lab{la} \\
b_p &=& (B\lambda_p+D_2)/g_pg_{p+1},
\lab{bp} \\
\Om_p \Om_{p-1}a_p^2 &=& (B\lambda_p+D_2)(B\lambda_{p-1}+D_2)/g_p^2
\nonumber \\
\quad &+& C_1\lambda_p\lambda_{p-1} +D_1(\lambda_p+\lambda_{p-1})-Q,
\lab{ap} \ea
where
\be
g_p=\lambda_p-\lambda_{p-1},\qquad \Om_p=\lambda_{p+1}-\lambda_{p-1}
\nonumber \ee

(ii) The analogous statement is valid for the dual basis:
\ba
K_2\phi_s &=& \mu_s\phi_s,
\lab{mu} \\
K_1\phi_s &=& c_{s+1}\phi_{s+1}+c_s\phi_{s-1}+d_s\phi_s,
\lab{dualt} \ea
where the spectrum $\mu_s$ and the matrix elements $c_s,d_s$ are obtained
from the expressions \re{la}, \re{bp}, \re{ap} by the substitutions:
\be
p\rar s, \lambda_p\rar\mu_s, a_p\rar c_s, b_p\rar d_s, C_{1,2}\rar C_{2,1},
D_{1,2}\rar D_{2,1}.
\lab{chng} \ee

The procedure \re{chng} expresses the symmetry property of the $AW(3)$ algebra.

Note that the discrete variables $p$ and $s$ are defined up to arbitrary
additive constant. However, if a $N+1$-dimensional representation of $AW(3)$
is considered then one has
\be
p=p_1+n, \qquad s=s_1+k,\quad n,k=0,1,...,N,
\lab{nk} \ee

with the obvious conditions
\be
a_{p_1}=d_{s_1}=a_{p_1+N+1}=d_{s_1+N+1}=0
\nonumber \ee

(iii) The overlaps between two eigenbases $\psi_p$ and $\phi_s$ are expressed
in terms of the Askey-Wilson polynomials \ci{AW1}
\be
\lphi \rpsi = W_k h_n
\; _4\Phi_3 \Biggl( {q^{-n},\; q^{-k},\; \beta q^{n-N}, \; \ga \de q^{k+1}\atop
q^{-N}, \;\beta \de q, \;\ga q}; q \Biggl| q \Biggr),
\lab{AS} \ee
where $W_k=\lphi \psi_{p_1} \rangle$ is the "vacuum amplitude", $h_n$ is
some normalization factor and $_4\Phi_3$ are the Askey-Wilson polynomials
(AWP) expressed in terms of basic hypergeometric function \ci{AW1}; the
parameters $\beta,\ga,\de$ of the AWP are expressed via the representations
parameters $B,C_{1,2},D_{1,2},Q$ of $AW(3)$ (for details see \ci{zh3}).

The formula $\re{AS}$ (found in \ci{zh3}) provides the solution of the Racah
problem because for the realization \re{K1}, \re{K2} the overlaps $\lphi \rpsi$
coincide with the Racah coefficients. Omitting the details of calculation,
we present the final result concerning the connection between the AWP and the
$sl_q(2)$ parameters:
\ba
N &=& \al_4-\al_3-\al_2-\al_1, \beta = -{w\over u}
q^{\al_1+\al_2+\al_4-\al_3-1},
\nonumber \\
\ga&=&{w\over v} q^{2\al_2-1},\qquad \de={z\over w} q^{2\al_3-1},
\nonumber \\
n&=&\al_{12}-\al_1-\al_2,\qquad k=\al_{23}-\al_2-\al_3
\lab{corpar} \ea

It may be verified that for the $su_q(2)$ and $su_q(1,1)$ algebras the
expressions
\re{AS} for the Racah coefficients coincide with those obtained in \ci{kir,
kl2,koo1,lisk}. However, the formula \re{AS} contains much more information
allowing one to obtain {\it all} classes of Askey-Wilson polynomials as
the Racah coefficients corresponding to addition {\it different} real forms
of $sl_q(2)$.

Indeed, the Racah coefficients for $su_q(2)$ and $su_q(1,1)$ algebras
correspond to the Askey-Wilson polynomials with $\cosh$-like spectra
$\lambda_p$ and $\mu_s$ , because in these cases $uw<0$ and $vz>0$ (see
\re{la} and \re{str}). The polynomials with other types of spectra can
be obtained only if the algebras with different structures are added.
Consider, for
example, the case $u=-v=w=z>0$. This leads to somewhat unusual (but fully
justified in our approach!) addition
\be
su_q(1,1)\oplus cu_q(2)\oplus su_q(2) = cu_q(2),
\lab{strng} \ee
involving three non-degenerate distinct types of $sl_q(2)$. Such an
addition rule has never been considered in the literature (note that
the addition
\re{strng} can not be obtained from the ordinary $su_q(2)$ or $su_q(1,1)$
additions
by renormalization of the generators with real parameters). Corresponding
intermediate Casimir operators have both $\sinh$-like spectrum. In the
"classical" limit $q\to 1$ the Racah coefficients for the addition \re{strng}
are expressed in terms of ordinary Krawtchouk polynomials, whereas the Racah
coefficients for $su_q(2)$ or $su_q(1,1)$ algebras become Racah polynomials.

It is clear that by choosing the other possible values for $u,v,w,z,$ we
exhaust
all the possible types of the Askey-Wilson polynomials in discrete variables
(in accordance with classification scheme of Ref.\ci{nik}). Indeed, there are
9 types of the polynomials corresponding to possible types
of the spectra $\lambda_p$ and $\mu_s$: $\cosh$-, $\sinh$- or $\exp$-.
These types correspond to 9 ways of combining the $(u,w)_{12}$ algebra
with the $(-v,z)_{23}$ one (with additional requirement that only the
representations $D_{\al_i}^+$ are admitted).

\section{Generalized Clebsch-Gordan problem and its hidden \break
symmetry}

In the previous Section we established the hidden symmetry algebra
$AW(3)$ underlying the Racah problem for $sl_q(2)$ algebra. In this Section
we show how one can obtain a new (generalized) Clebsch-Gordan problem
and the corresponding
Clebsch-Gordan coefficients (GCGC) from the Racah scheme by means of some
simple
contraction procedure.

Consider again the Racah scheme \re{K1}, \re{K2} and suppose that $u\ge 0$ (if
$u=0$ we suppose in addition that $v\ge0$). This condition means that the
algebra $(u,v)_1$ has the representation $D_{\al_1}^+$, so the spectrum
$n+\al_1$ is unbounded ($n=0,1,...,\infty$). Consider the limit $n\to\infty$.
It is easily verified that the operators $\Xi_+=\exp(-\om A_0^{(1)})A_+^{(1)}$
and
$\Xi_-=A_-^{(1)}\exp(-\om A_0^{(1)})$ commute with one another in this limit
and hence can be replaced by constants:
\be
\Xi_+\to\xi, \qquad \Xi_-\to\xi^\ast,
\lab{const} \ee
where
\be
|\xi|^2=uq/(1-q)
\lab{xi} \ee
(note that other physical implications of the degeneration \re{const}
of the
$sl_q(2)$ algebra were studied in \ci{slava} ).

The value  $\ka_1$ of the Casimir operator
$\hat\ka_1$ remains a real constant \re{vcas} in this limit.

Then the algebra $(u,v)_1$ is completely degenerate and the operator $K_1$
\re{K1} becomes in this limit
\be
K_1=\xi A_-^{(2)}\exp(\om A_0^{(2)}) + \xi^\ast \exp(\om A_0^{(2)}) A_+^{(2)}
+\ka_1\exp(2\om A_0^{(2)}).
\lab{K1l} \ee

The operator $K_2$ \re{K2} remains obviously unchanged, and the full Casimir
operator \re{full} becomes
\ba
\hat\ka_4 = K_1\exp(2\om A_0^{(3)}) + \xi A_-^{(3)}\exp(\om A_0^{(3)}) +
\xi^\ast\exp(\om A_0^{(3)})A_+^{(3)}
\nonumber \\
= \xi A_-^{(23)}\exp(\om A_0^{(23)}) + \xi^\ast\exp(\om A_0^{(23)})A_+^{(23)}
+ \ka_1\exp(2\om A_0^{(23)})
\lab{ka4} \ea

Thus we get a generalized Clebsch-Gordan problem for two $sl_q(2)$ algebras
($i=2,3$), because diagonalization of the operator $K_1$ \re{K1l} corresponds
to
the choosing of some "twisted" eigenbasis $\psi_p$ for the representation space
of the algebra $(-v,w)_2$, whereas diagonalization of the operator $K_2$
corresponds to choosing the connected basis $\phi_s$ on the representation
space of the algebra $(u,w)_{12}$. Then the operator $\hat\ka_4$ \re{ka4}
is exact analogue of the operator $K_1$ for the algebra $(u,w)_{12}$ and
plays the role of the projection of the total momentum (in terminology of
the ordinary Clebsch-Gordan problem), obviously commuting with $K_1$ and
$K_2$.

It is clear, by the construction, that $AW(3)$ remains to be a hidden
symmetry algebra underlying the generalized Clebsch-Gordan problem for
$sl_q(2)$.

The corresponding GCGC are obtained from the Racah coefficients by using the
procedure \re{const}, \re{xi}: we again get the expression in terms of Askey-
Wilson polynomials \re{AS}  where the parameters are given by the expressions
\re{corpar}. By the way, we automatically obtain the explicit expression  for
the spectrum \re{ka12} of the operator $K_1$ \re{K1l}.

The "standard" Clebsch-Gordan problem for $sl_q(2)$ is obtained if we put
$u=0$. Then $\xi=0$ and the operator $K_1$ \re{K1l} is reduced to one term
\be
K_1=\ka_1 \exp(2\om A_0^{(2)}).
\lab{simple} \ee

It is clear that the diagonalization of the operator \re{simple} correponds to
choosing the canonical basis $|n;\al_2\rangle$ \re{rep}. Then the
overlaps between the eigenstates of the operators $K_1$ and $K_2$ are nothing
else than ordinary Clebsch-Gordan coefficients for adding two different
types of $sl_q(2)$ algebra. In this case one of the AWP parameters tends to
infinity: $\beta\to\infty$ \re{corpar} and the basic function $_4\Phi_3$
is reduced to
$_3\Phi_2$ (as is easily seen from \re{AS}) and we get expression of CGC in
terms of the Hahn q-polynomials. For the $su_q(2)$ and $su_q(1,1)$ algebras
this result is well known \ci{kir,vak,kl1,koo1,lisk}. However for
adding of different types of $sl_q(2)$ this result is new (in another form
it was presented in Ref. \ci{grkh}).

What is the "classical" $(q=1)$ analogue of the GCGC? It is seen from \re{K1l}
that for $q\to 1$ the operator $K_1$ becomes (up to a constant term) the
simple linear combination of the Lie algebra's generators:
\be
K_1 = \xi a_-^{(2)} + \xi^\ast a_+^{(2)} + \eta a_0^{(2)} + const,
\lab{limm} \ee
where $\eta = \lim _ {q\to 1} {(-2\om \kappa_1)}$ and $a^{(2)}$ are the
generators of the corresponding Lie algebra.

The diagonalization of the operator $K_1$ \re{limm} corresponds to choosing
of "rotated" basis $\psi_p$ in the space of $D_{\al_2}^+$, i.e.
$\psi_p = U|p;\al_2\rangle$, where U is a unitary automorphism of corresponding
Lie group (say, rotation for $su(2)$). Then in the "classical" limit the
GCGC coincide with ordinary CGC because unitary automorphisms of Lie algebra
can not destroy the CGC. However, for $q\ne 1$ the operator $K_1$ \re{K1l}
can not be obtained from \re{simple} by means of any unitary transformation
because these operators have essentially different spectra. So in a q-domain
GCGC do not coincide with standard CGC and the Clebsch-Gordan problem
for $sl_q(2)$ becomes non-trivial and essentially depends on the choice of
appropriate basis. Perhaps, that is why the GCGC were not found earlier. Note,
that "twisted" GCGC for $su_q(2)$ algebra were calculated (by means of
$AW(3)$ symmetry) in Ref. \ci{twi}.

It is worth mentioning that the contraction procedure \re{const} takes place
only in q-domain and has no classical analogue. So for the q-case the
transition from Racah problem to generalized Clebsch-Gordan one is a  much
simpler procedure than in the "classical" case (it is instructive
to compare \re{const} with limiting
procedure allowing to get CGC from Racah ones for ordinary $su(2)$ \ci{mosk}
and for $su_q(2)$ \ci{kl2}; that procedure is quite different from the
considered here).
We plan to discuss this strange phenomenon elsewhere.

\section{Conclusion}
We have shown that the same $AW(3)$ algebra serves as a hidden symmetry
underlying both the Racah and generalized Clebsch-Gordan problems. The $AW(3)$
algebra provides very simple way to obtain explicit expressions for the
corresponding coefficients in terms of the
Askey-Wilson polynomials. Most properties
 of  these coefficients (symmetry, generating functions, recurrent relations)
can be automatically derived from representations of $AW(3)$ (see, e.g. Ref.
\ci{grz1} where similar analysis was applied to explain the symmetry properties
of $6j$-symbols of the ordinary $su(2)$ algebra).

In this paper we restricted ourselves to the representations of discrete series
$D_\al^+$. However the formulae \re{AW3}, \re{str} describing the realization
of $AW(3)$ algebra are equally valid for all the possible representation
series. So one
can obtain the expressions for Racah and Clebsch-Gordan coefficients connecting
the representations of different series (negative discrete $D_\al^-$,
principal $C$ etc.): then we obtain the Askey-Wilson polynomials with
continuous or mixed spectrum. This will be considered separately.

\section{Aknowledgements}
One of the authors (A.Zh.) is very grateful to L.Vinet and
V.Spiridonov for helpful discussions and hospitality in the Laboratory of
Nuclear Physics of the University of Montreal.

\end{document}